\documentclass[a4paper]{spie}  

\usepackage{amsmath,amsfonts,amssymb}
\usepackage{graphicx}
\usepackage[colorlinks=true, allcolors=blue]{hyperref}
\usepackage{textgreek} 
\usepackage{subcaption}
\usepackage{rotating}

\title{A Fourier optics approach to evaluate the astrometric performance of MICADO}

\author[a,b,c]{J.A. van den Born}
\author[b,d]{W. Jellema}
\author[a]{R. Navarro}
\author[b]{E. Tolstoy}
\author[c]{B. Jayawardhana}
\author[a]{A.W. Janssen}
\affil[a]{\footnotesize NOVA Optical/IR Instrumentation Group, Oude Hoogeveensedijk 4, 7991 PD Dwingeloo, The Netherlands}
\affil[b]{\footnotesize Kapteyn Astronomical Institute, PO Box 800, 9700 AV Groningen, The Netherlands}
\affil[c]{\footnotesize Engineering and Technology Institute Groningen, Nijenborgh 4, 9747 AG Groningen, The Netherlands}
\affil[d]{\footnotesize SRON Netherlands Insitute for Space Research, PO Box 800, 9700 AV Groningen, The Netherlands}

\authorinfo{E-mail: born@astro.rug.nl}

\begin{document} 
\maketitle

\begin{abstract}
We present our investigation into the impact of wavefront errors on high accuracy astrometry using Fourier Optics. MICADO, the upcoming near-IR imaging instrument for the Extremely Large Telescope, will offer capabilities for relative astrometry with an accuracy of 50~micro arcseconds (\textmu as). Due to the large size of the point spread function (PSF) compared to the astrometric requirement, the detailed shape and position of the PSF on the detector must be well understood. Furthermore, because the atmospheric dispersion corrector of MICADO is a moving component within an otherwise mostly static instrument, it might not be sufficient to perform a simple pre-observation calibration. Therefore, we have built a Fourier Optics framework, allowing us to evaluate the small changes in the centroid position of the PSF as a function of wavefront error. For a complete evaluation, we model both the low order surface form errors, using Zernike polynomials, and the mid- and high-spatial frequencies, using Power Spectral Density analysis. 
The described work will then make it possible, performing full diffractive beam propagation, to assess the expected astrometric performance of MICADO.
\end{abstract}


\section{INTRODUCTION}
\label{sec:intro}  

MICADO, the Multi-AO Imaging CamerA for Deep Observations, is one of three first generation instruments in development for the Extremely Large Telescope (ELT)\cite{Davies2018}. One of the primary use cases for MICADO will be using the unprecedented resolution of the telescope to perform highly precise astrometric measurements, complementary to the Gaia catalog. The instrument will be able to determine the relative distance between multiple point sources to an accuracy of 50 microarcseconds (\textmu as)\cite{Pott2018}. The main challenge for reaching such performance lies in the understanding of the system stability - how image distortions change as the telescopes moves around, temperature variations occur and time passes. In this work we investigate the impact of wavefront errors on the geometric distortions, and subsequently on the astrometric accuracy. We do this by investigating the small shifts in the centroid location due to surface errors on the MICADO ADC optics. Sources in different parts of the field have different footprints on the prisms, and will thus experience slightly different wavefront perturbations. The shape of the point spread function at multiple points in the field of view will therefore differ. We will investigate this effect.

In the next section, we will summarize the Fourier Optics theory that we used. Section~\ref{sec:paraxial_micado_model} details the simplified optical system used for our analyses. Then, Sec.~\ref{sec:results} and Sec.~\ref{sec:discussion} will show and discuss how the surface errors on the ADC optics will influence the astrometric performance of the MICADO instrument.

\section{FUNDAMENTAL EQUATIONS OF THE SIMULATION FRAMEWORK}
\label{sec:theory}
\subsection{Diffractive Beam Propagation Using Fourier Theory}\label{subsec:fourierTheory}
The propagation of light through an optical system, taking into account the diffractive properties of its wavelike nature, can be done using scalar diffraction theory. The two methods we used in this work are the angular spectrum method for plane waves and the Fresnel diffraction integral\cite{Goodman2005}. 

In collimated space, where wavefronts are expected to be flat, the angular spectrum method can be used to calculate the propagation of a wavefront over a distance $z$. First, the spectrum $A_{\textnormal{in}}$ of the known complex input field, $U_{\textnormal{in}}(x,y)$ is calculated. The output field, $U_{\textnormal{out}}(x,y)$, at a distance $z$ from the input plane is calculated as
\begin{align}
    U_{\textnormal{out}}(x,y)&=\iint \displaylimits_{-\infty}^{\phantom{abc}\infty} A_{\textnormal{in}} \left( u, v \right) \exp{ \left[i k_z z \right] } \exp{ \left[ i 2 \pi \left( u x + v y \right) \right] } du dv \nonumber\\
    &= \mathcal{F}^{-1}\left\{ A_{\textnormal{in}} \left(u, v \right) \exp{ \left[i k_z z\right] }\right\}, \label{eq:angular_spectrum_method}
\end{align}
where the frequencies $u=k_x/(2\pi)$ and $v=k_y/(2\pi)$ implicitly define the value of $k_z=2\pi\sqrt{\lambda^{-2}-u^2-v^2}$. The three $\mathbf{k}$-space components follow from the wave vector
\begin{equation}
   \mathbf{k} = \begin{pmatrix} k_x \\ k_y \\ k_z \end{pmatrix},
\end{equation}
with magnitude $|\mathbf{k}|= k =2\pi/\lambda$.

The Fresnel diffraction integral describes the propagation of a curved wavefront from a plane with coordinates $(\xi, \eta)$ to another plane with coordinates $(x,y)$. The output plane is located a distance $z$ away from the input plane. Given an input field distribution $U_{\textnormal{in}}(\xi, \eta)$, the output field distribution, $U_{\textnormal{out}}(x,y)$, is described by 
\begin{equation}
    U_{\textnormal{out}}(x,y) = \frac{e^{ikz}}{i\lambda z} e^{\frac{ik}{2z}(x^2+y^2)} \iint \displaylimits_{-\infty}^{\phantom{abc}\infty} \bigg\{ U_{\textnormal{in}}(\xi, \eta) e^{\frac{ik}{2z}(\xi^2 + \eta^2)}\bigg\} e^{-i \frac{2\pi}{\lambda z}(x\xi + y\eta)}d\xi d\eta. \label{eq:FresnelApprox}
\end{equation}

This expression resembles a Fourier transform evaluated at frequencies $\tilde{u}=\frac{x}{\lambda z}$ and $\tilde{v}=\frac{y}{\lambda z}$, where we have used a tilde to denote that these frequencies are not equivalent to the frequencies $u$ and $v$ of the angular spectrum method.

When one would be interested in analysing the properties of an off-axis point source, a phase tilt can be added to the input field, resulting in a shift of the central peak of the diffraction pattern as described by the shift theorem. If $\mathcal{F}\left\{ g(x,y)\right\} = G(u,v)$, then the shift theorem states that
\begin{equation}
    G(u-u_o, v-v_o) = \mathcal{F}\left\{g(x,y)e^{i2\pi(x u_o + y v_o)}\right\}.
\end{equation}

This property is particularly useful when performing off-axis beam propagations. Since numerical calculations of the frequency response are not over an infinite plane we can carefully choose the calculation region around $u_o$ and $v_o$ to make sure that the response of the optical system is properly represented.

Assume we have a system that brings a uniformly illuminated aperture with radius $r$ to a focus at a distance $z$. The optical system can be summarized as a single phase transformer to an input field $U_{\textnormal{in}}(\xi,\eta)$. To find the central peak intensity at the focal distance of the mirror at coordinates $(x_o, y_o)$ we add a phase tilt
\begin{equation}
    \begin{aligned}
        \phi(\xi, \eta) &= k\Delta z(\xi, \eta)\\
        &= \frac{2 \pi}{\lambda}\left( \xi \tan \theta_{\xi} + \eta \tan \theta_{\eta} \right),
    \end{aligned}
\end{equation}
or in exponential form
\begin{align}
    U_{\textnormal{in}}(\xi,\eta) = e^{i \phi} &= e^{i\frac{2\pi}{\lambda}\left( \xi \tan \theta_{\xi} + \eta \tan \theta_{\eta} \right)}.\label{eq:phase_tilt}
\end{align}
In the above, $\theta_{\xi}$ and $\theta_{\eta}$ are the field angles in the $\xi$ and $\eta$ directions, respectively.

We noted earlier that the Fresnel integral is a Fourier transformation from the spatial coordinates $(\xi, \eta)$ to the frequency coordinates $(\tilde{u}, \tilde{v}) = (x/\lambda z, y/\lambda z)$. We define a new set of coordinates, such that $\hat{x} = x - x_o$ and $\hat{y} = y - y_o$. This results in a new translated set of frequency coordinates
\begin{equation}
    \begin{aligned}
        \hat{u} = \frac{\hat{x}}{\lambda z} = \frac{x - x_o}{\lambda z},\\
        \hat{v} = \frac{\hat{y}}{\lambda z} = \frac{y - y_o}{\lambda z}.
    \end{aligned}
\end{equation}

Considering the above, we recognize that the phase tilt we introduced results in a shift of the image given by
\begin{equation}
    \begin{aligned}
        x_o &= z \tan \theta_{\xi},\\
        y_o &= z \tan \theta_{\eta}.
    \end{aligned}
\end{equation}

For off-axis beam propagation in collimated space, the beam will be tilted with respect to the optical axis. We can apply similar reasoning as above to the angular spectrum method for plane waves to show that the shift $(u_o,v_o)$ of the peak in the frequency domain is described by 
\begin{equation}
    \begin{aligned}
        u_o &= \tan \theta_{x}/\lambda,\\
        v_o &= \tan \theta_{y}/\lambda.
    \end{aligned}
\end{equation}
This frequency shift changes the value of $k_z$ in Eq.~\eqref{eq:angular_spectrum_method}, making the amplitude shift less apparent, but nonetheless straight forward to calculate.

\subsection{Describing Wavefront Errors With Power Spectral Density Analysis}

We follow Refs.~\citenum{Sidick2009} and \citenum{Agocs2018} to model the wavefront errors (WFE). We use Zernike polynomials\cite{mahajan2013} for the low order shape errors and a Lorentzian Power Spectral Density (PSD) profile for the mid and high spatial frequency content of the wavefront error. The Lorentzian PSD distribution described in detail in Ref.~\citenum{Sidick2009} has shown to accurately describe measured mirror characteristics and allows us to define a WFE by only three parameters. These are the root mean square (RMS) value of the WFE $\sigma$, the power slope coefficient $p$ and the cut-off frequency $\rho_c$. The PSD is given by
\begin{equation}
\textnormal{PSD}(u_m,v_n) = \frac{\sigma^2 A}{h_0}\frac{1}{1+(\rho_{mn}/\rho_{c})^{p}},
\label{eq:SidickPSD}
\end{equation}
with the normalization factor
\begin{equation}
h_0 = \sum^{M}_{m=1}\sum^{N}_{n=1}\frac{1}{1+(\rho_{mn}/\rho_{c})^{p}}
\end{equation}
and the radial frequency
\begin{equation}
\rho_{mn} = \sqrt{u_m^2 + v_n^2},
\end{equation}
where $u_m$ and $v_n$ are the discrete spatial frequencies and $A$ is the area of the surface for which the PSD curve is defined. The area under the PSD curve is equal to $\sigma^2$, due to the normalization of Eq.~\eqref{eq:SidickPSD} by $h_0$.

To arrive at a surface profile $h(x,y)$ that can be used as a wavefront error in our simulations, a random uniformly distributed phase profile $\phi(u,v)$ is added and the inverse Fourier Transform is used, resulting in the expression
\begin{equation}
h(x,y) 
= \mathcal{F}^{-1} \left\{ e^{i\phi(u,v)} \sqrt{A \times \textnormal{PSD}(u,v)} \right\}. \label{eq:inversePSD}
\end{equation}

\section{Model of the ELT-MICADO system}
\label{sec:paraxial_micado_model}
\begin{figure}
    \centering
    \includegraphics[width=0.7\textwidth]{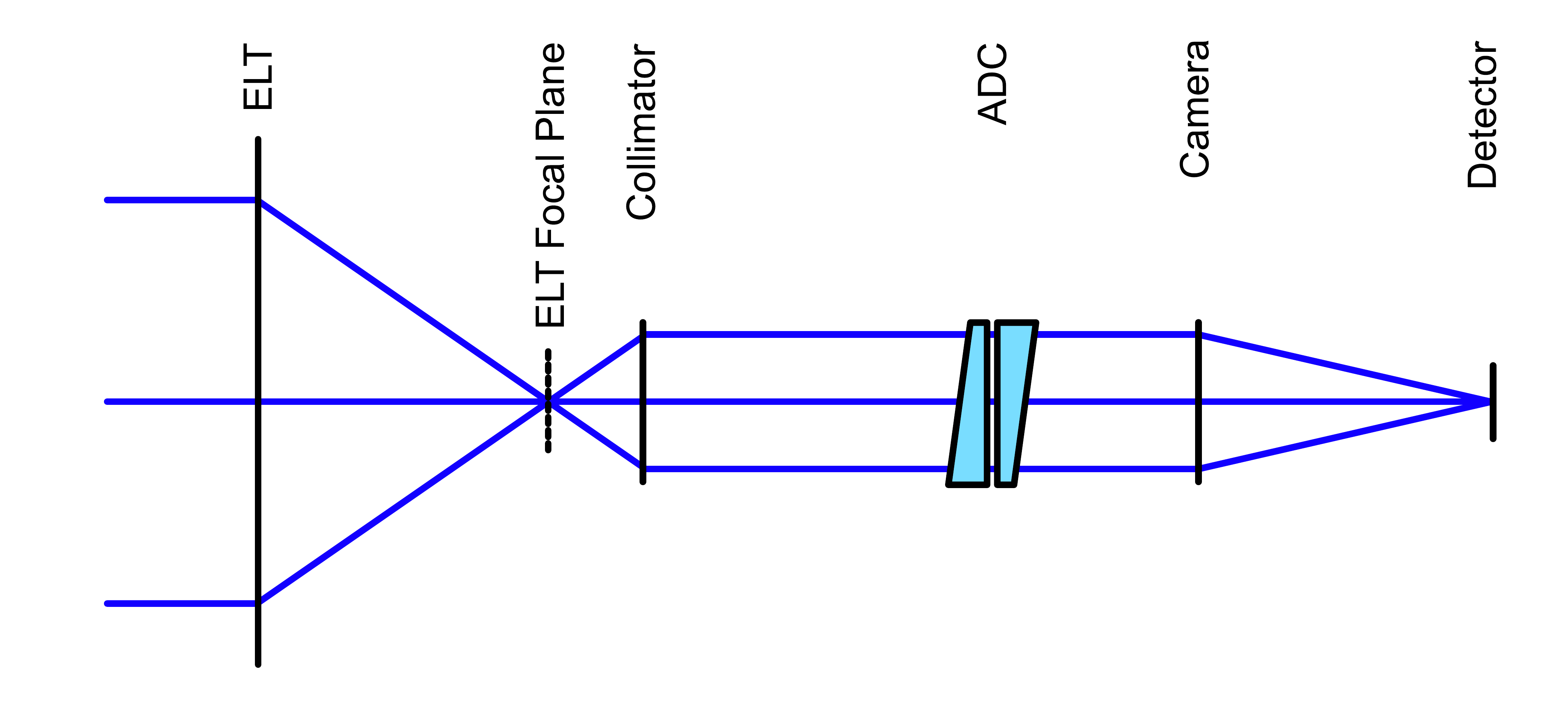}
    \caption{The paraxial model that we use to model the ELT-MICADO system.}
    \label{fig:ELT-MICADO_paraxial}
\end{figure}

\begin{figure}
\centering
\begin{minipage}[t]{.47\textwidth}
\centering
\captionsetup{width=0.9\linewidth}
\includegraphics[width=\linewidth]{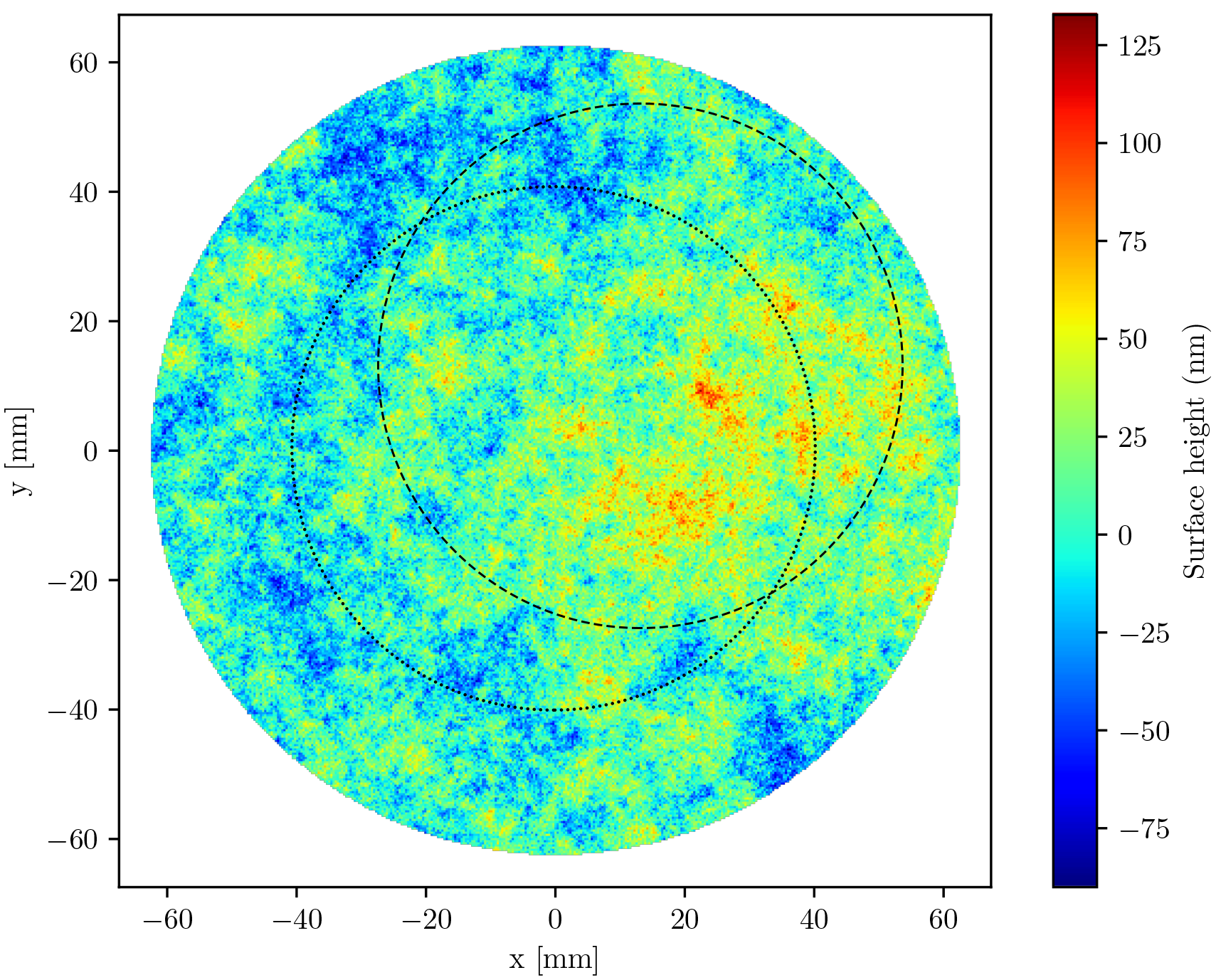}
\captionof{figure}{An instance of a generated surface profile used on the first ADC prism. The wavefront error was generated from a PSD with $p=2$, $\rho_c=1$ and $\sigma=25$~nm. The beam footprint is represented by a dotted line for an on-axis beam and by a dashed line in the corner of the field.}
\label{fig:adc_wfe}
\end{minipage}%
\hfill
\begin{minipage}[t]{.47\textwidth}
\centering
\captionsetup{width=0.9\linewidth}
\includegraphics[width=\linewidth]{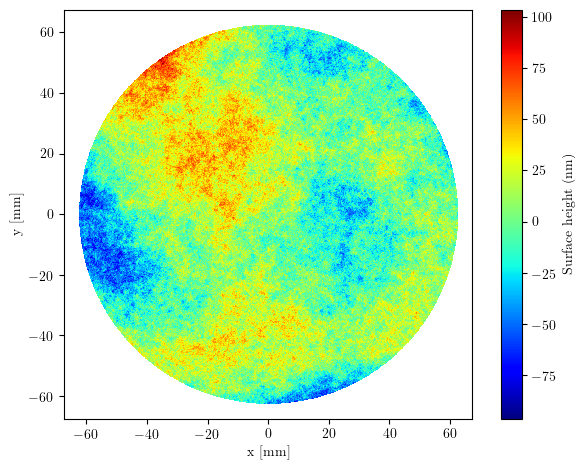}
\captionof{figure}{Another instance of a generated surface profile. Here $\sigma_{\textnormal{\tiny{total}}}=25$~nm. The low order shape errors are determined by Zernike polynomials ($\sigma_{\textnormal{\tiny{Zernike}}}=20$~nm) and the mid and high spatial frequency content is given by a PSD ($\sigma_{\textnormal{\tiny{PSD}}}=15$~nm, $p=2$ and $\rho_c=1$).}
\label{fig:adc_wfe_psd_zern}
\end{minipage}%

\end{figure}

We have modelled a paraxial model of the ELT-MICADO system as shown in Fig.~\ref{fig:ELT-MICADO_paraxial}. First, the light from the sky gets focused by the 39 meter primary mirror of the ELT. Some distance after the first focal plane, the beam gets collimated by the MICADO collimator mirrors. In the collimated space the beam passes through the atmospheric dispersion corrector (ADC) and finally gets focussed onto the detector by the camera mirrors. The various mirrors and the ADC are implemented in the beam propagation as simple phase transformations. The amplitude and phase distribution of an on-axis beam are shown in Appendix~\ref{sec:appendixA} for several locations in the system. Surface profile perturbations have been modelled for all four prisms of the MICADO Atmospheric Dispersion Corrector and are applied as phase perturbations. Figure~\ref{fig:adc_wfe} shows an example of a surface profile defined by a PSD curve only and Fig.~\ref{fig:adc_wfe_psd_zern} shows an example of a surface profile defined by a combination of both Zernike modes and a PSD curve. 

Lab measurements of high quality optics reveal that the power slope coefficient $p$ typically ranges from 1.3 to 3, depending on the manufacturing and polishing processes used\cite{Alcock2010,Duparre2002,Toebben1996,Walsh1999}. Therefore, we will use $p=2$ and $\rho_c=1$ for a representative PSD curve in this work.

MICADO will have a field of view of roughly 54"$\times$54", corresponding to a focal plane area of roughly 550~mm by 550~mm in this paraxial system. In reality, the detector area will be approximately 200 mm by 200 mm. To model the image response over such a wide field we use the principles described in Sec.~\ref{subsec:fourierTheory}, where we use ray transfer matrices to find the ray angles within the system. 

The analyses in the next section are done at a wavelength of 1.49 \textmu m, where the radius of the first zero of the PSF is at 0.09 mm from the PSF core, corresponding to a PSF of 9 mas. The platescale at the focal plane is 100.13~mas/mm. The beam propagation has been done at a numerical resolution of 1001$\times$1001 and an initial sampling resolution of 160 mm per resolution element. The analysis of Sec.~\ref{sec:geometric_distortions} has been performed at an increased numerical resolution of 3001$\times$3001 and initial sampling resolution of 50 mm per resolution element. The sampling resolution of the subsequent surfaces depends on the output field coordinates of the propagation up to that surface.
Finally, we assume an obscured entrance aperture with an outer diameter of 38.5 meters and and inner diameter of 10.7 meter, resulting in an obscuration ratio $\epsilon = 0.278$.

 \section{Results}
 \label{sec:results}
 \subsection{Impact of Wavefront Errors on the On-axis Centroid Position}
Traditionally, the impact of wavefront perturbations on the Strehl ratio, $S$, is given by the Mar\'echal approximation,
\begin{equation}
    S(\sigma) = \exp\left[-\left(\frac{2\pi\sigma}{\lambda}\right)^2\right]. \label{eq:marechal}
\end{equation}

This expression works best when the root mean square value (RMS) of the wavefront phase perturbation, $\sigma$ is relatively small and contains mostly high spatial frequencies\cite{Ross2009}. Therefore, we expect that a wavefront error described by a PSD curve follows Eq.~\eqref{eq:marechal} more closely, than one described by a limited number of Zernike modes. 

\begin{figure}
\centering
\begin{minipage}[t]{.47\textwidth}
\centering
\captionsetup{width=0.9\linewidth}
\includegraphics[width=\linewidth]{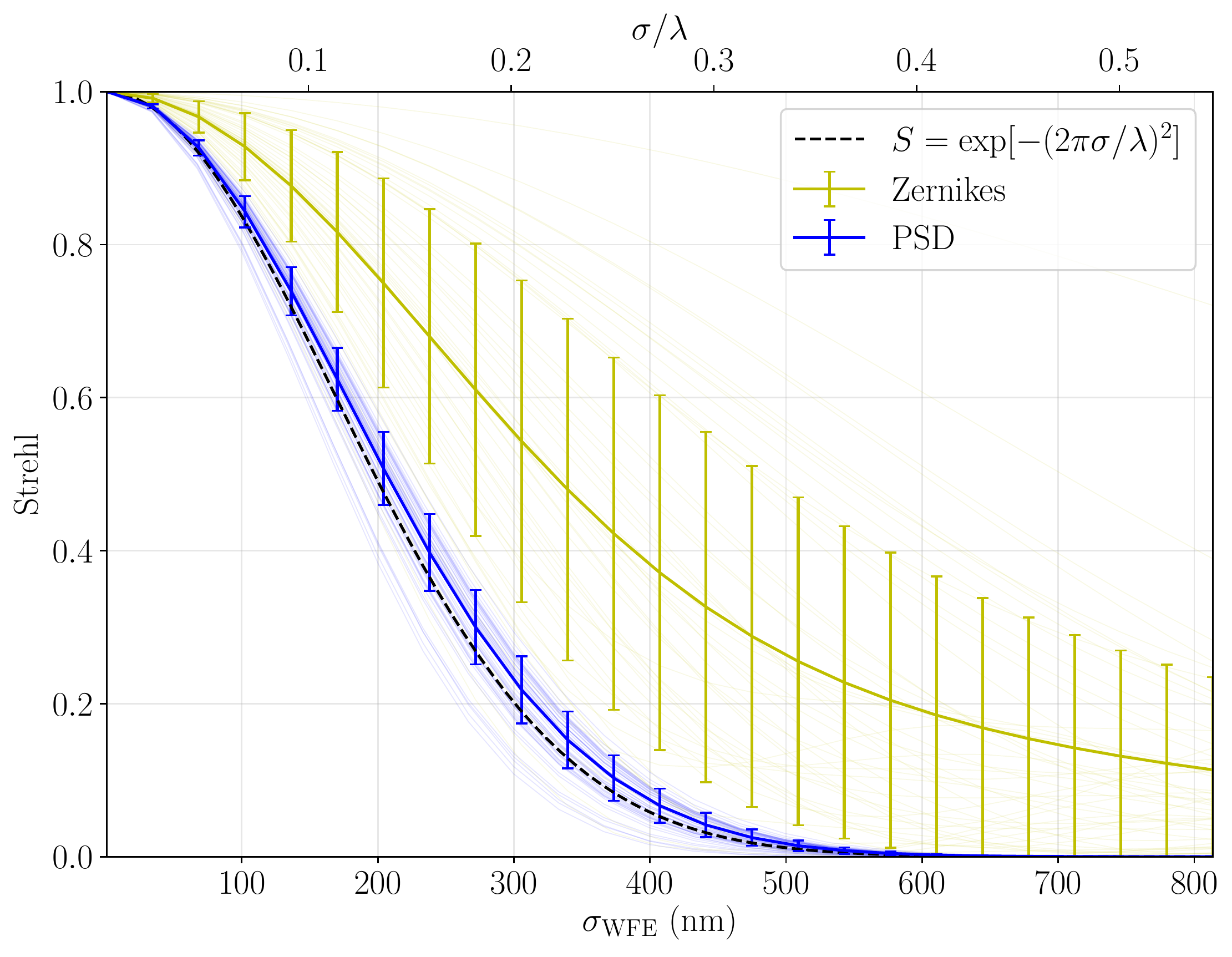}
\captionof{figure}{Relationship between the RMS of the wavefront error and the Strehl ratio, for 50 different instances of a WFE applied to the first ADC prism defined by the same PSD. The bright lines show the mean and standard deviations of the 50 iterations.}
\label{fig:strehl_relation}
\end{minipage}%
\hfill
\begin{minipage}[t]{.47\textwidth}
\centering
\captionsetup{width=0.9\linewidth}
\includegraphics[width=\linewidth]{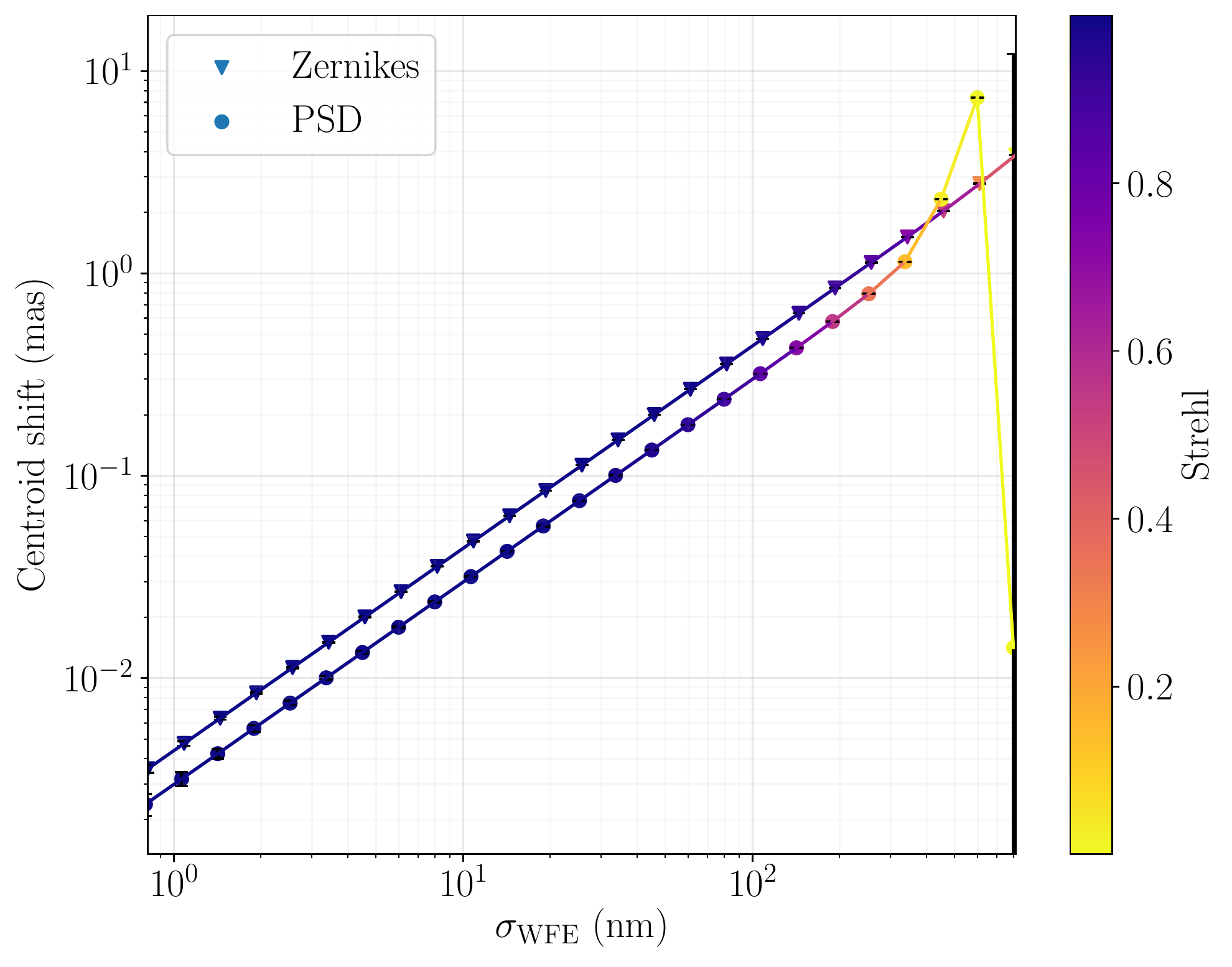}
\captionof{figure}{The centroid shift in milli arcseconds as a function of the RMS value of the WFE, for both Zernike representation (yellow) and PSD representation (blue). For extremely large wavefront perturbations, the PSF falls apart and the fit of the centroid fails.}
\label{fig:wfe_scaling_relation}
\end{minipage}
\end{figure}

As a verification of our simulation, we reproduce the Maréchal relation. First, we add a surface profile to the first ADC prism, defined by a PSD with power slope $p=2$ and cut-off frequency $\rho_c=1$. By taking the phase information from one particular instance of the surface profile generated by this PSD, we can scale the RMS of this surface profile to a desired different value, without changing any other property. Then, we fit 66 Zernike coefficients to this instance of the surface profile to find a similar surface form error in Zernike mode representation. Because the exact shape of a particular instance of a surface profile is random, we repeat this exercise 50 times. The impact on the Strehl ratio is then examined, as shown in Fig.~\ref{fig:strehl_relation}. We have scaled the surface profile RMS by $(n-1)$ to find the RMS of the phase perturbation, or WFE. As expected, we see that the WFEs defined by the PSD curve follow the Mar\'echal approximation much more closely than the WFEs defined by the 66 Zernike polynomial coefficients. The large variation in the Strehl ratio over the different iterations using Zernike modes, can be explained as a limitation to how accurately the generated WFE can be fit by only 66 Zernike modes. As more coefficients are fitted, the Zernike representation should approach the Mar\'echal approximation more closely and more consistently.

Using this same scaling procedure of the $\sigma_{\textnormal{WFE}}$, we now inspect the centroid shift of the PSF by fitting an obscured Airy pattern. The intensity of the PSF at coordinates $(x,y)$ is given as
\begin{equation}
    I(x,y) = I_{bg} + I_0  \left[ \frac{2J_1\left( r\right)}{r} - \frac{2 \epsilon J_1\left( \epsilon r\right)}{{r}} \right]^2,
\end{equation}
where
\begin{align}
    r &= \frac{\pi R_z}{R} \sqrt{(x-x_0)^2 + (y-y_0)^2},\nonumber\\
    R_z &= 1.2196698912665045...,\nonumber\\
    J_1(x) &= \sum^{\infty}_{m=0}\frac{(-1)^m}{m!(m+1)!}\left(\frac{x}{2}\right)^{2m+1}.\nonumber
\end{align}
Here $I_{bg}$ is the background intensity, $I_0$ is the peak intensity, $\epsilon$ is the obscuration ratio, $R$ is the radius of the first zero and $J_1(x)$ is the first order Bessel function of the first kind. The centroid is located at $(x_0, y_0)$.

Figure~\ref{fig:wfe_scaling_relation} shows a linear scaling relation between the wavefront error RMS and the centroid shift in milliarcseconds for an on-axis PSF. Different instances of the generated surface profile give a slightly different y-amplitude, but the slope is constant. This results makes sense if it is interpretated as an increasing tip and tilt component of the WFE over the beam footprint. Even at large $\sigma$, the average tip and tilt over the 80 mm diameter of the beam at the ADC still falls well within the small angle limit, where you expect a linear relation between the centroid shift on the focal plane and the tip and tilt of the WFE at the ADC.

\begin{figure}
    \centering
    \includegraphics[width=0.6\textwidth]{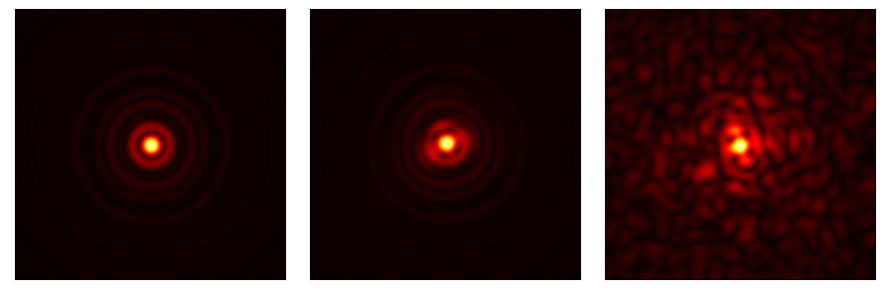}
    \caption{The left panel shows the PSF amplitude without any applied perturbations. The middle panel shows the PSF when a significant WFE is added to the first ADC prism, described by 66 Zernike polynomials. The right panel shows the PSF when a WFE described by a PSD with the same $\sigma$ as in the middle panel is added.}
    \label{fig:perturbed_PSF}
\end{figure}

When the phase perturbations are of significant magnitude the point spread function will disintegrate, as illustrated in Fig.~\ref{fig:perturbed_PSF}. Then, it might not be preferable to use an (obscured) Airy pattern to fit the centroid location. A simple barycenter fit, Moffat or Gaussian profile or overlap integral could then be considered\cite{Moffat1969,Goldsmith1998}.

\begin{figure}
\centering
\begin{minipage}[t]{.47\textwidth}
\centering
\captionsetup{width=0.9\linewidth}
\includegraphics[width=\linewidth]{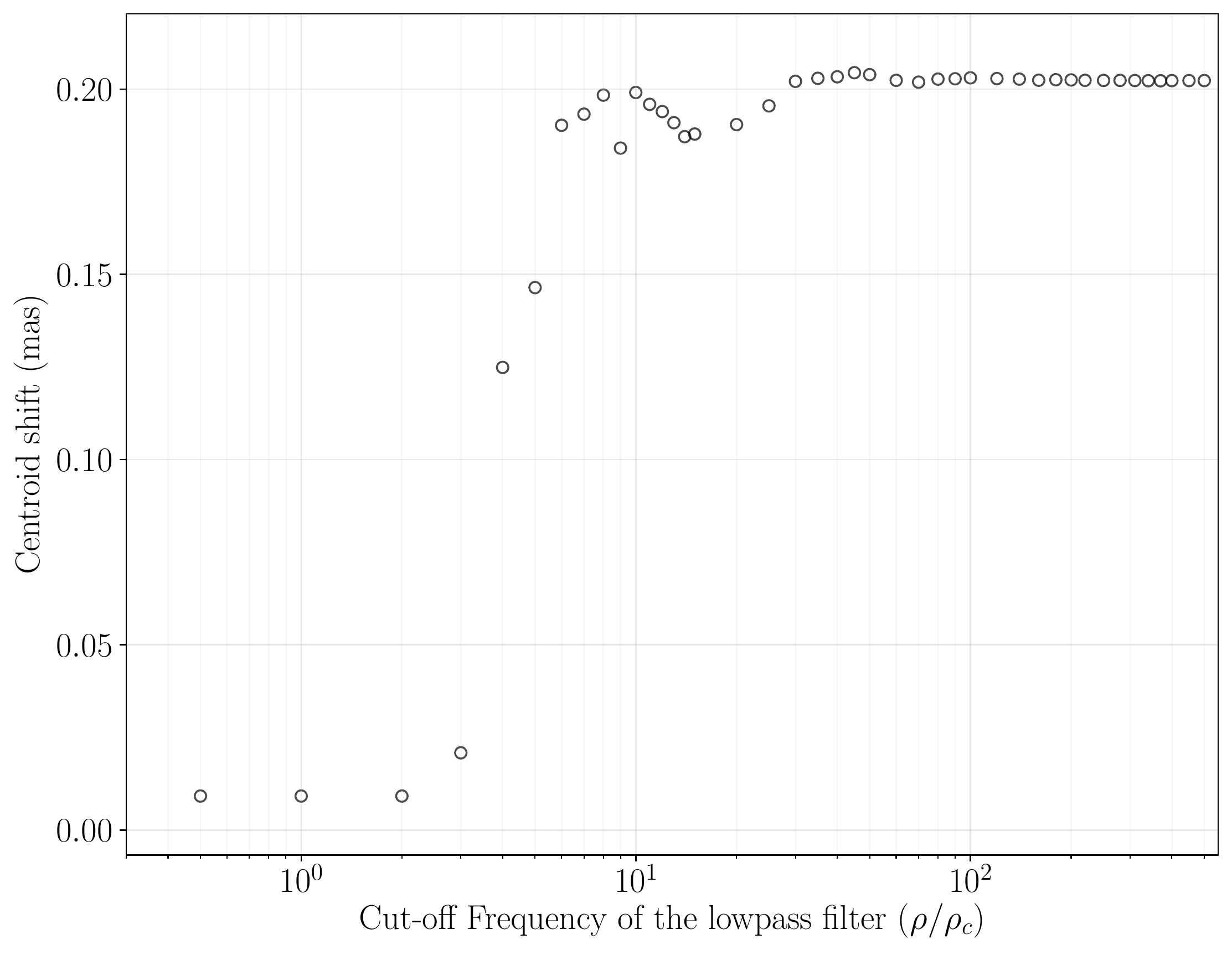}
\captionof{figure}{The centroid shift of the PSF as a function of the cut-off frequency of a lowpass spatial filter on the WFE. Most of the centroid shift is reached after roughly 10 cylces/aperture.}
\label{fig:bandpass}
\end{minipage}%
\hfill
\begin{minipage}[t]{.47\textwidth}
\centering
\captionsetup{width=0.9\linewidth}
\includegraphics[width=\linewidth]{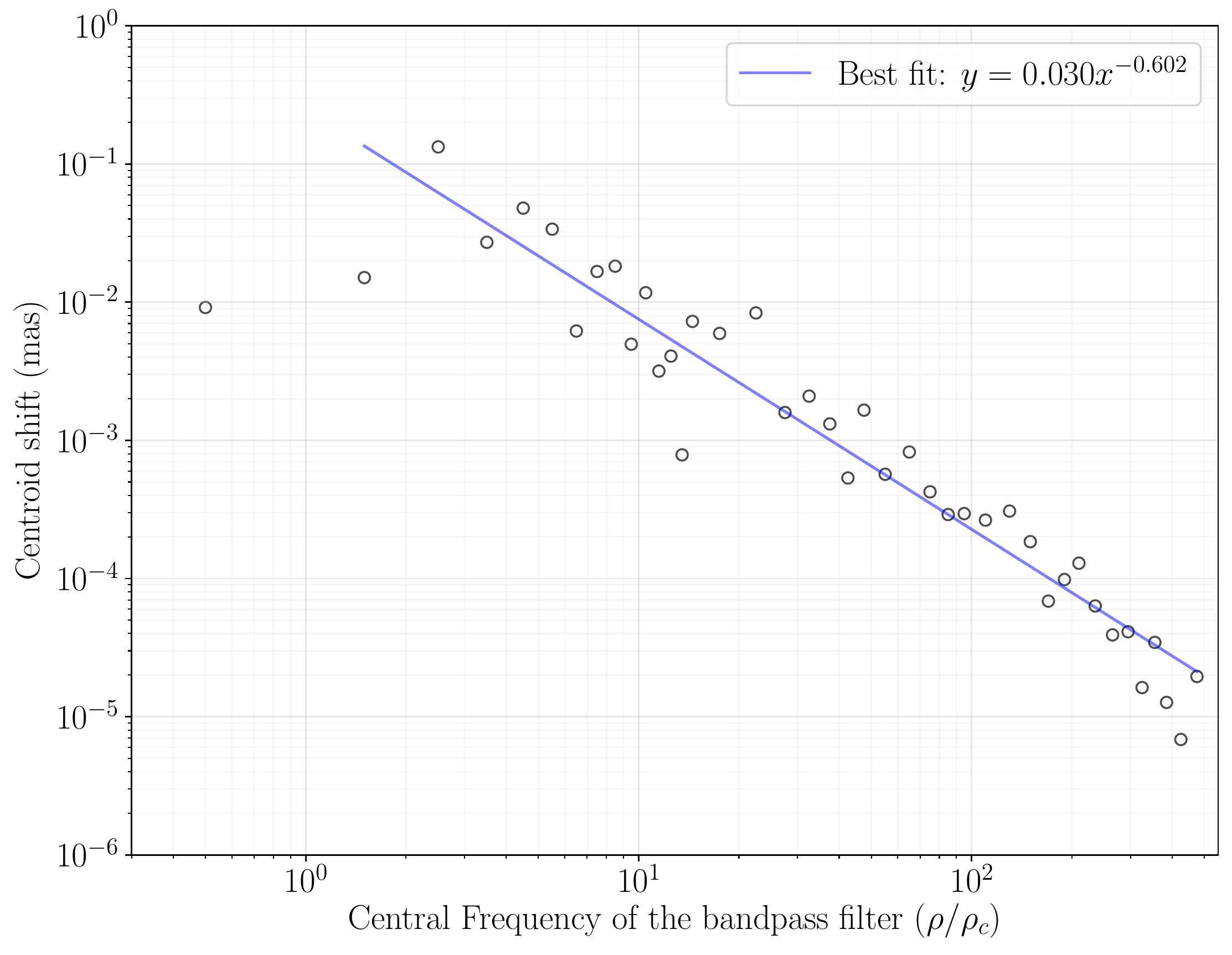}
\captionof{figure}{The centroid shift of the PSF as a function of the spatial frequency. An exponentially decreasing relation can be observed.}
\label{fig:lowpass}
\end{minipage}
\end{figure}

By filtering the spatial frequency content of the applied surface profile we can inspect the contributions of the increasing spatial frequencies to the overall centroid shift. Figure~\ref{fig:bandpass} shows the cumulative centroid shift for one particular instance of a surface profile as we incrementally increase the upper limit of the included spatial frequency band. The high spatial frequencies contribute little to the overall shift. This is confirmed by Fig.~\ref{fig:lowpass}, where we have inspected small spatial frequency bandpasses of this surface profile separately. An exponential decrease is seen as a function of the spatial frequency. A few instances of the filtered WFE and their effect on the PSF shapes are given in Appendix~\ref{sec:appendixB}. In practice, this will mean that only up to a spatial frequency of approximately 10 cycles/aperture will contribute appreciably to the astrometric performance of the instrument, given that the PSD cut-off frequency $\rho_c$ is small.

\subsection{Wavefront Error Induced Geometric Distortions}\label{sec:geometric_distortions}
Optical distortions in the MICADO instrument do not necessarily need to be minimized to micro arcsecond levels for optimum astrometric performance, but they should be stable over long timescales. This concept concerns the MICADO ADC in particular, as it is one of the components that can move during an observation. In the previous section, we have seen that surface perturbations introduce a shift in the centroid location. We also know that beams corresponding to different field angles will have different footprints on the ADC surfaces, even though the ADC is located close to, but not exactly at the exit pupil. Therefore, the centroid shift should differ as a function of field position.

\begin{figure}
    \centering
    \includegraphics[width=0.6\linewidth]{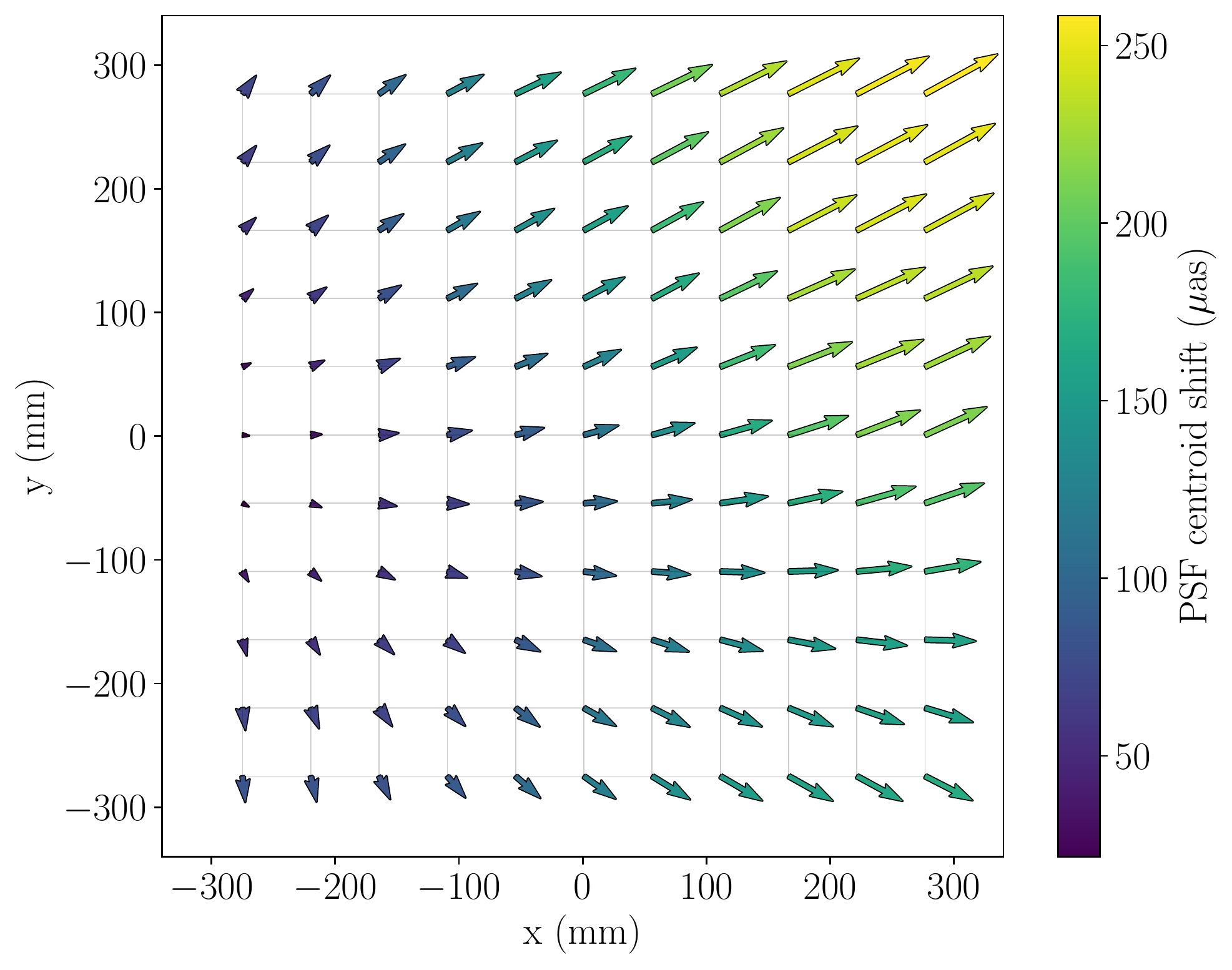}
    \caption{The change in the centroid location over the full field, before and after applying a WFE to all ADC prisms, defined by a PSD ($p=2$, $\rho_c=1$ and $\sigma=25$~nm). The RMS of the PSF centroid shift is 146 \textmu as.}
    \label{fig:GD_PSD_only}
\end{figure}

\begin{figure}
    \centering
    \includegraphics[width=0.6\linewidth]{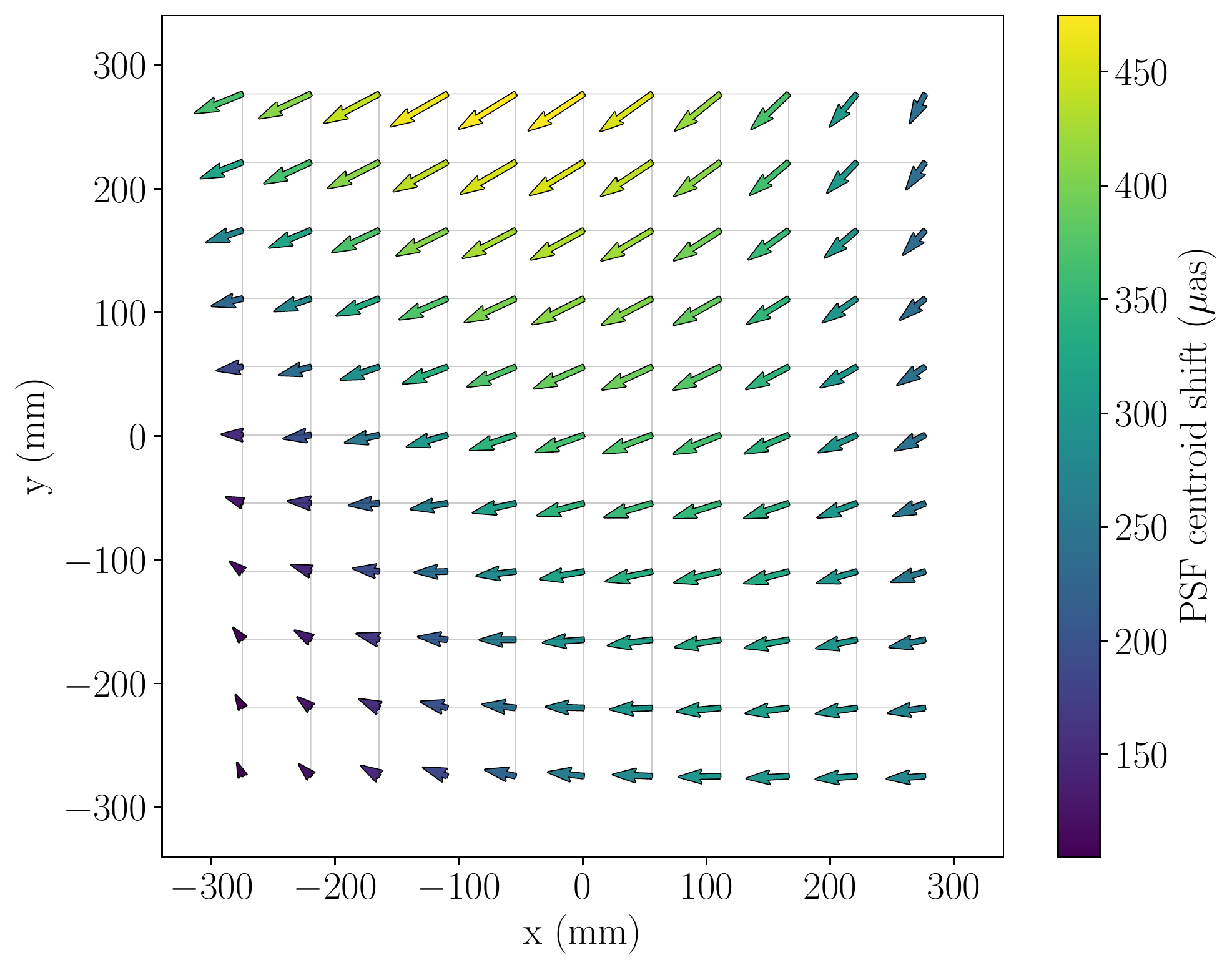}
    \caption{The change in the centroid location over the full field, before and after applying a WFE to all ADC prisms. Each of the four WFEs is defined by lower order form errors, described by Zernike polynomials, and a PSD for the mid and high spatial frequencies. Here $\sigma_{\textnormal{Zernike}}=20$ nm and $\sigma_{\textnormal{PSD}}=15$ nm, giving $\sigma_{\textnormal{WFE}}=25$ nm. The RMS of the PSF centroid shift is 316 \textmu as.}
    \label{fig:GD_PSD_Zern}
\end{figure}

We have investigated this idea by applying representative wavefront errors to each of the four ADC prisms. Then, we performed the beam propagation for a grid of $11\times11$ field points. Each resulting PSF was compared to the non-perturbed version at that same position and fitted to an Airy pattern to measure the centroid shift. The results are shown in Fig.~\ref{fig:GD_PSD_only} for WFEs defined by a PSD only and in Fig.~\ref{fig:GD_PSD_Zern} for WFEs defined by both Zernike modes and a PSD curve. The total RMS of the WFE is kept the same in both simulations.

Variations of the PSF centroid shift over the field are apparent without any clear symmetry. The transition from one region to the next is smooth and continuous. The centroid shift is at most several hundred \textmu as. The exact shape of the distortions depends very much on the exact form of the WFE, as illustrated by the differences in the distortions of Figs.~\ref{fig:GD_PSD_only} and \ref{fig:GD_PSD_Zern}. At 316 \textmu as RMS centroid shift, the WFEs described by both Zernike and PSD contributions show larger shifts, compared to the WFEs described by a PSD curve only, which shows a RMS shift of 146 \textmu as. While the sample size here is too small to draw any firm conclusions, these results do suggest that, indeed, the low order spatial distributions dominate the total distortion magnitude.

 \section{Discussion}
 \label{sec:discussion}
 Modern astrometry has in large part been possibly by an increasing understanding of the instrumental distortions and the stability thereof\cite{Anderson2003,Bellini2011,Pott2018,Rodeghiero2019,Service2016,Trippe2010}. The distortions shown in the previous section are relatively small, compared to the typical geometric distortions expected from the optical design and telescope stability. Furhermore, the shape of these distortion patterns can be resolved by a third order polynomial\cite{Trippe2010}, resulting in an astrometric solution accurate to 8 \textmu as, with a minor improvement when using a fifth order polynomial. This solution is well within the astrometric requirement of the instrument. However, because the ADC is able to rotate during observations it is useful to understand the change in the distortion for a given amount of rotation of the ADC prisms. While astrometric calibrations will be performed regularly on MICADO\cite{Rodeghiero2019}, the astrometric uncertainty will increase as the ADC moves away from the calibrated configuration. One solution to this problem is not to rotate the ADC at all during an exposure. Under the proper conditions this might indeed be the best way forward. Further research is necessary to investigate this in more detail.

The Fourier optics approach presented in this work is a computationally expensive method to calculate the centroid shift of the PSF. We suspect, but have not yet confirmed, that most of this PSF centroid shift can be calculated from the average tilt on the footprint of the plane that introduces the WFE. If true, the wavefront error induced geometric distortions should vary more severely for surfaces close to the system focal plane, while the distortions close to the pupil plane are more controlled. 

During the manufacturing of the ADC prism blanks and the subsequent polishing, the accuracy with which the (transmitted) wavefront errors or surface errors are measured will likely be limited in resolution. As shown, it will not be necessary to measure the complete PSD curve to the level of micro-roughness to get an accurate grasp on the introduced distortions, but an effort should be made to measure the Mid Spatial Frequency errors on the surfaces. We will further investigate this in future work and derive quantitative calibration requirements as part of the development of an ADC calibration strategy.
 
 \section{Conclusions}
 \label{sec:conclusions}
 In this work we have outlined the impact that wavefront errors have on the centroid positions of the MICADO point spread function over the full instrumental field of view. A rigorous Fourier Optics approach was chosen to fully take into account the diffractive nature of light, allowing us to inspect the small contributions of low, mid and high spatial frequencies of the wavefront error on the centroid position. We have shown that most of the PSF centroid shift is caused by spatial frequencies of the WFE up to approximately 10 cycles/aperture.

MICADO aims to provide relative astrometric accuracy on the order of several tens of micro arcseconds. Significant effort to understand the instrument distortions and temporal stability is being made to ensure that this level of performance will be achieved. The geometric distortions induced by the polishing imperfections on the ADC prisms will contribute to the distortions on the order of several tenths of a milli arcsecond. We believe these distortions will not be a major concern for the astrometric performance as it is currently foreseen that the ADC will not be rotating during an astrometric exposure and can therefore be calibrated out with a pre-observation calibration.
 
\bibliography{report} 
\bibliographystyle{spiebib} 

\appendix
\label{sec:appendix}
\section{On-axis beam propagation of the ELT-MICADO model}\label{sec:appendixA}
\rotatebox{90}{
\begin{minipage}{0.97\textheight}
    \begin{minipage}[t]{0.45\textwidth}
        \centering
        \includegraphics[width=\textwidth]{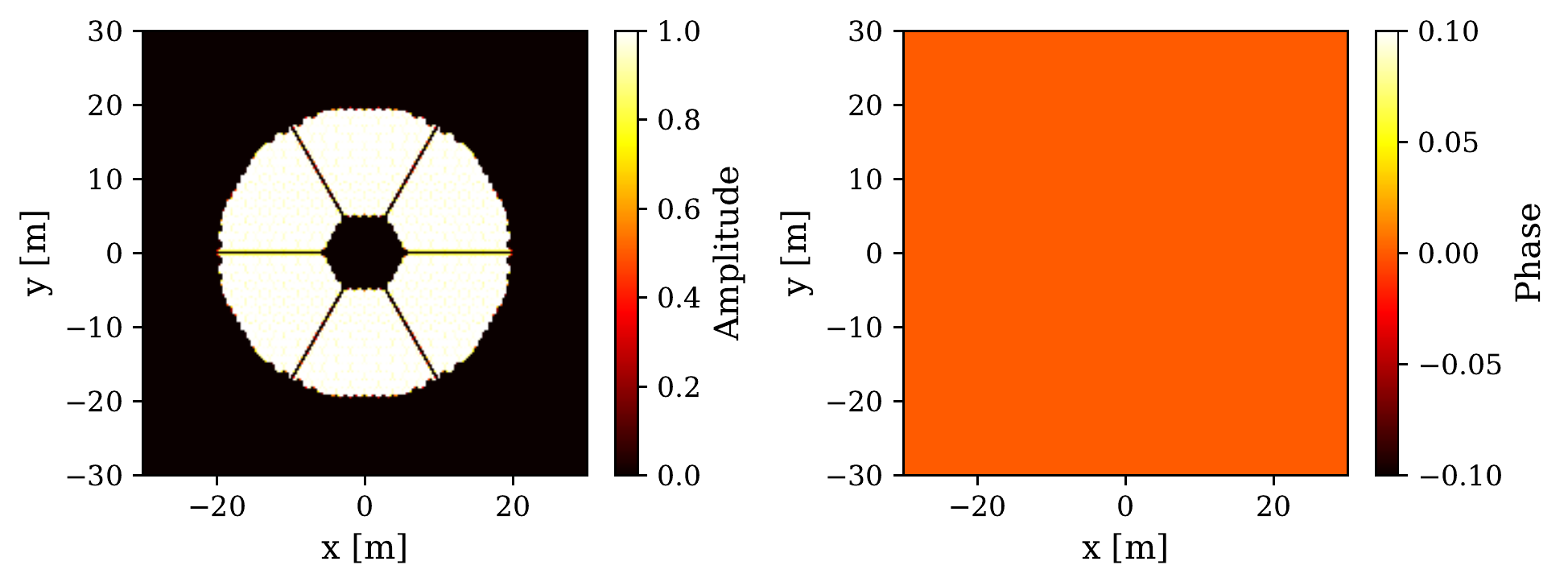}
        \label{fig:fourier_ELT-M1}
    \end{minipage}
    \hspace{0.05\textwidth}
    \begin{minipage}[t]{0.45\textwidth}
        \centering
        \includegraphics[width=\textwidth]{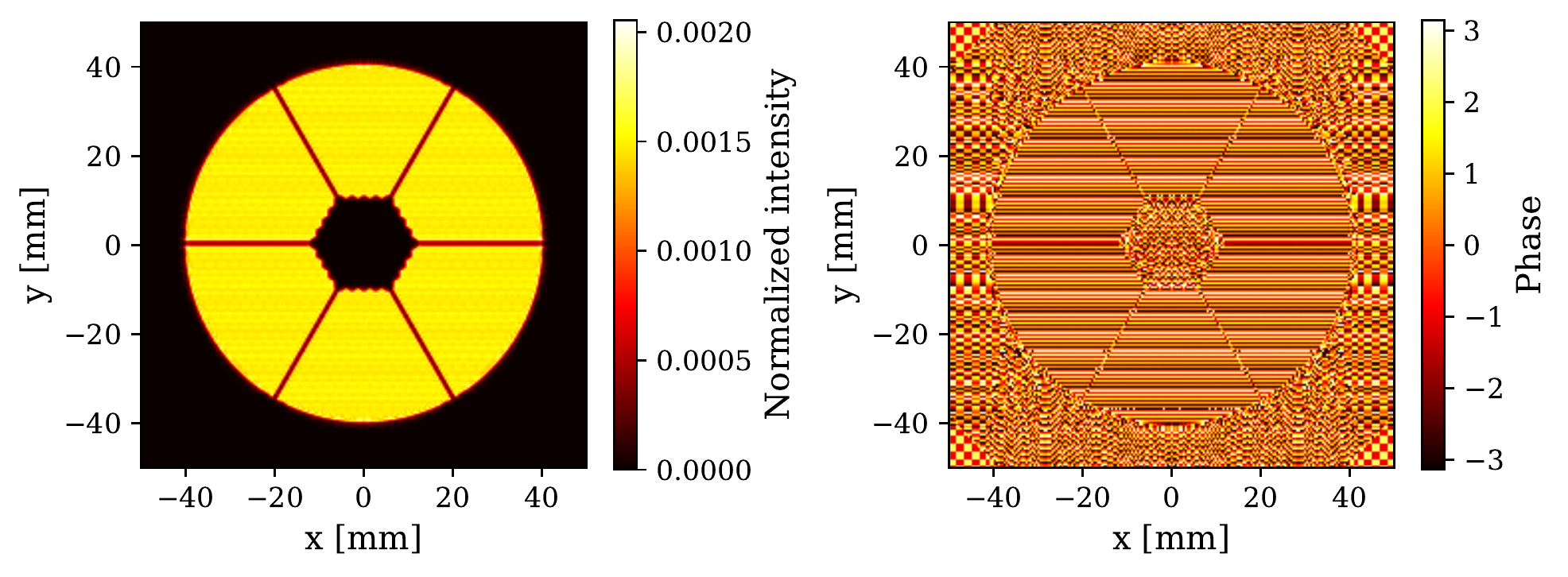}
        \label{fig:fourier_ELT-MCD-ADC-P2}
    \end{minipage}
    \\
    \begin{minipage}[t]{0.45\textwidth}
        \centering
        \includegraphics[width=\textwidth]{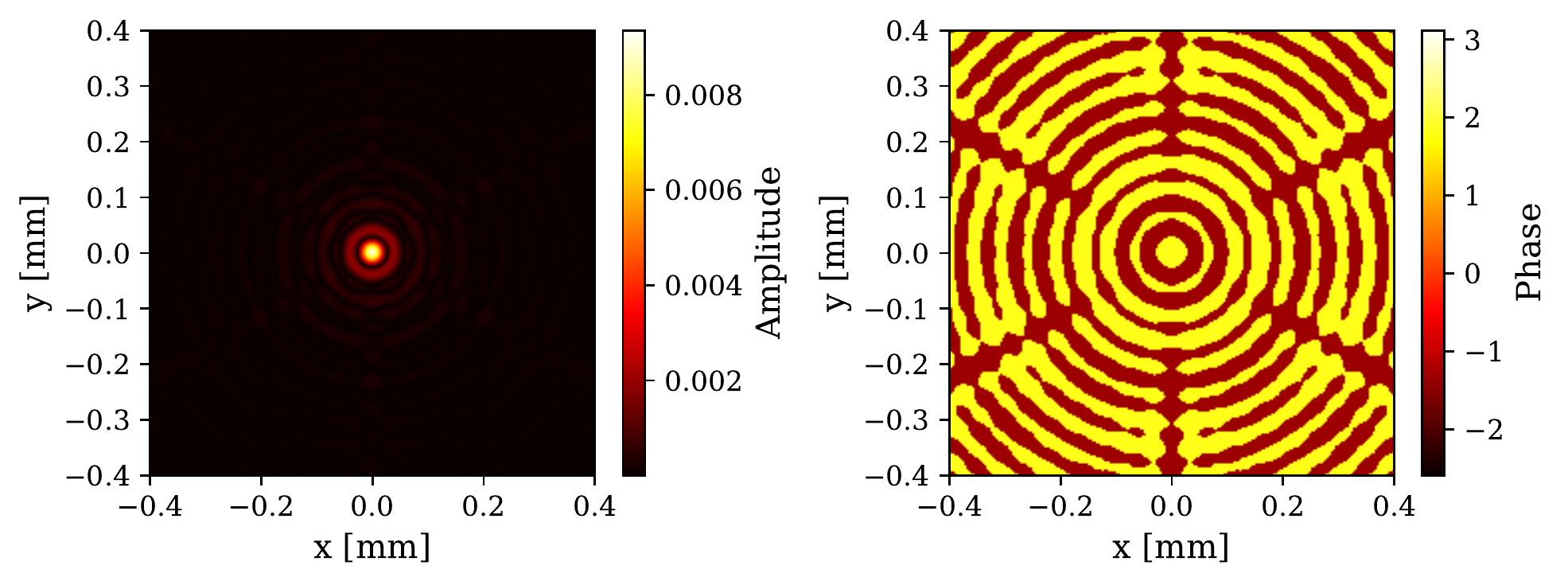}
        \label{fig:fourier_ELT-IFP}
    \end{minipage}  
    \hspace{0.05\textwidth}
    \begin{minipage}[t]{0.45\textwidth}
        \centering
        \includegraphics[width=\textwidth]{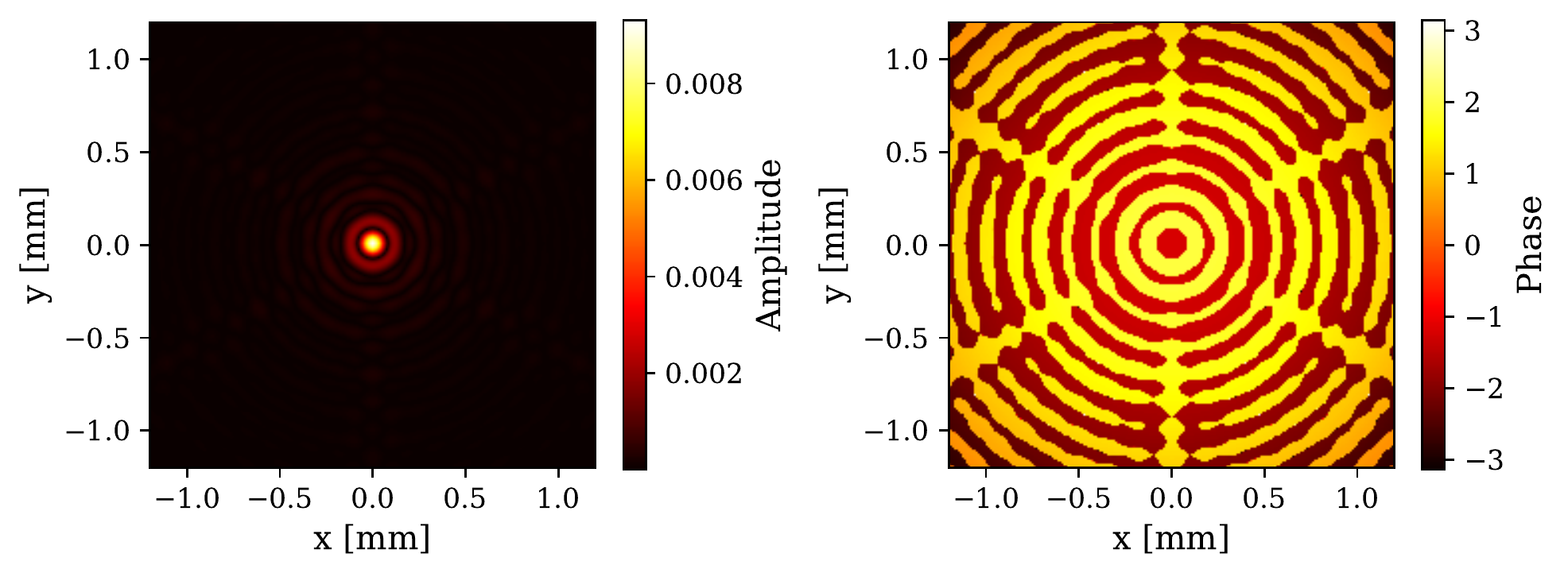}
        \label{fig:fourier_ELT-MCD-DET}
    \end{minipage}
    \\
    \begin{minipage}[t]{0.45\textwidth}
        \centering
        \includegraphics[width=\textwidth]{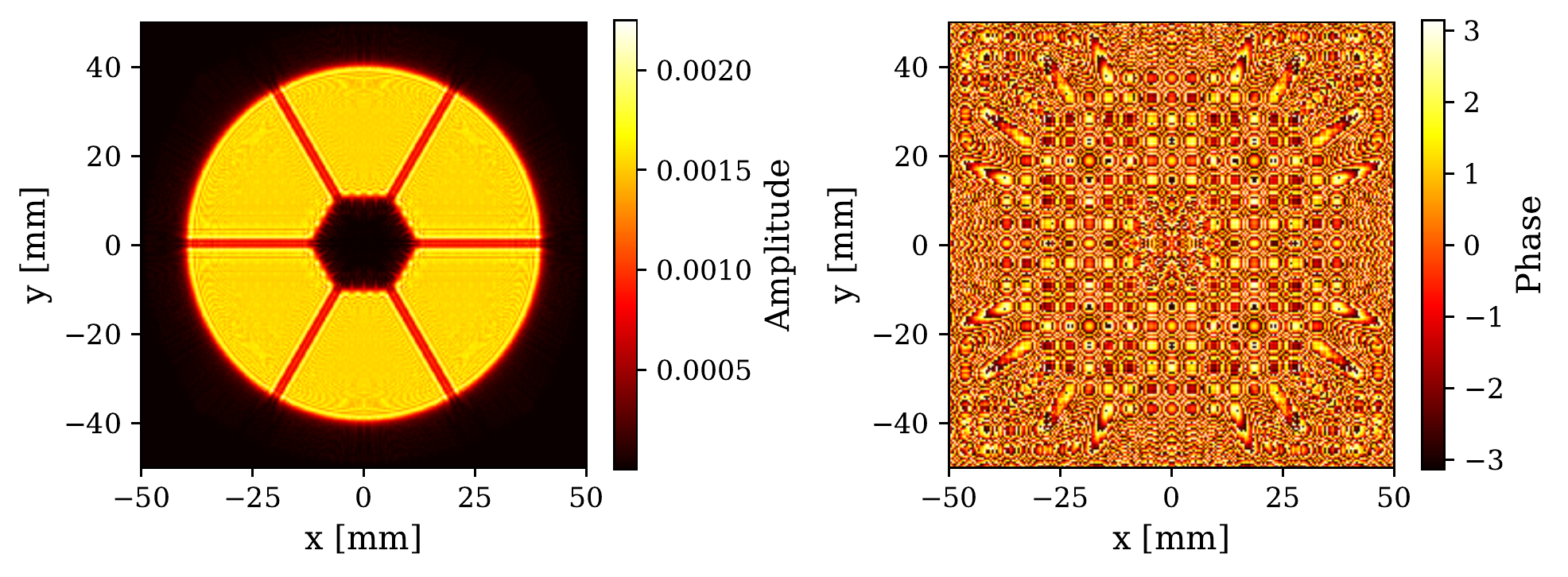}
        \label{fig:fourier_ELT-COL}
    \end{minipage}%
    \captionsetup{width=.9\textheight}
    \captionof{figure}{The amplitude and phase distribution at the different surfaces of the ELT-MICADO paraxial model. Top left: ELT-M1, middle left: ELT focal plane, bottom left: MICADO collimator, top right: between the two ADC amici prisms, bottom right: MICADO focal plane.\\These figures were generated at a wavelength of 1.49 \textmu m and at a resolution of $3001\times3001$ with an input pixel scale of 50 mm per resolution element.}\label{fig:fourier_ELT-MICADO}
\end{minipage}%
}

\section{Spatially Filtered Wavefront Errors}\label{sec:appendixB}
\subsection{Lowpass Filtered Wavefront Errors}

\begin{figure}[h!]
    \begin{subfigure}[t]{\textwidth}
        \centering
        \includegraphics[width=\textwidth]{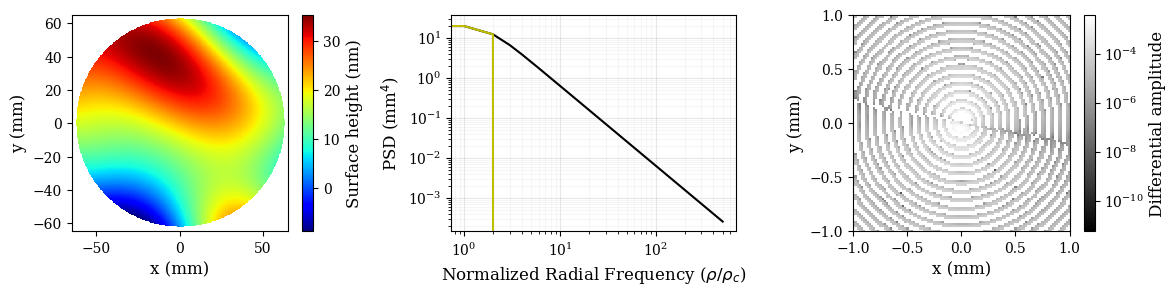}
        \label{fig:lowpass_1}
    \end{subfigure}\vspace{0.5cm}
    \\
    \begin{subfigure}[t]{\textwidth}
        \centering
        \includegraphics[width=\textwidth]{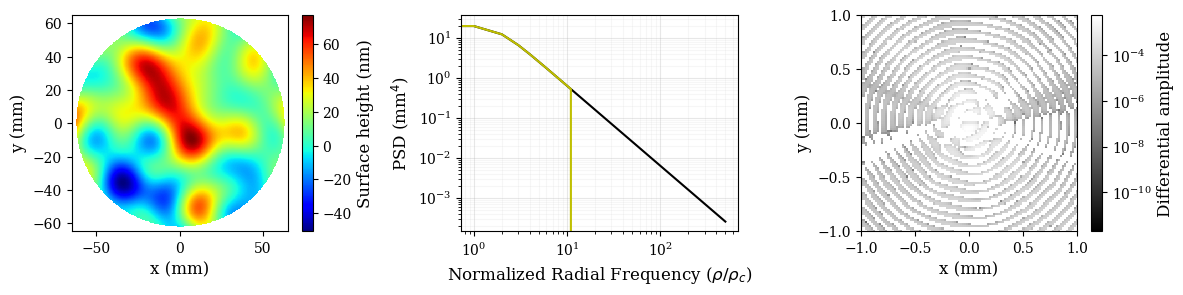}
        \label{fig:lowpass_2}
    \end{subfigure}\vspace{0.5cm}
    \\
    \begin{subfigure}[t]{\textwidth}
        \centering
        \includegraphics[width=\textwidth]{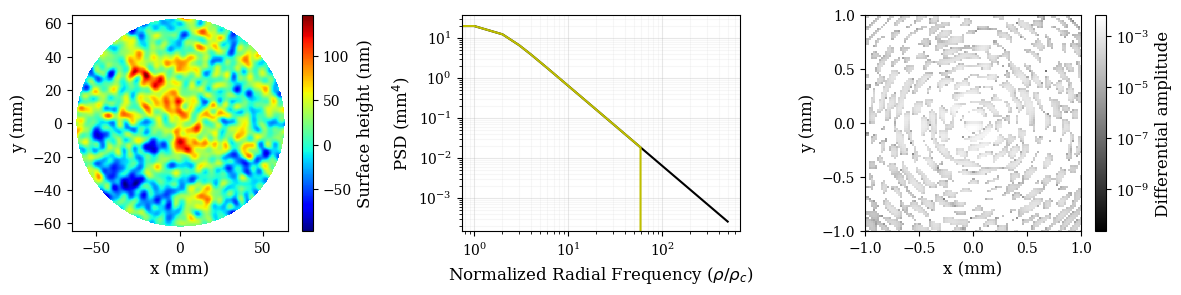}
        \label{fig:lowpass_3}
    \end{subfigure}
    \caption{Three instances of the lowpass filtered WFE. As the cut-off frequency of the lowpass filter increases the original WFE appears more faithfully. The left panel shows the spatially filtered wavefront error on the ADC surface. The middle panel shows the part of the PSD curve that is used for the filtered WFE (orange line) and the original full PSD curve (black line). Finally, the right panel shows the differential amplitude between the unperturbed PSF and the PSF where the WFE in the first panel has been applied at the ADC.}
    \label{fig:lowpass_multifigure}
\end{figure}

\newpage
\subsection{Bandpass Filtered Wavefront Errors}
\begin{figure}[h!]
    \begin{subfigure}[t]{\textwidth}
        \centering
        \includegraphics[width=\textwidth]{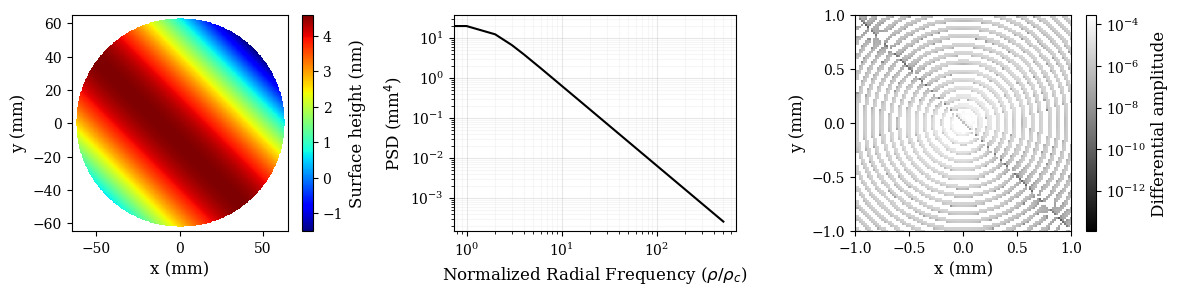}
        \label{fig:bandpass_1}
    \end{subfigure}\vspace{1cm}
    \\
    \begin{subfigure}[t]{\textwidth}
        \centering
        \includegraphics[width=\textwidth]{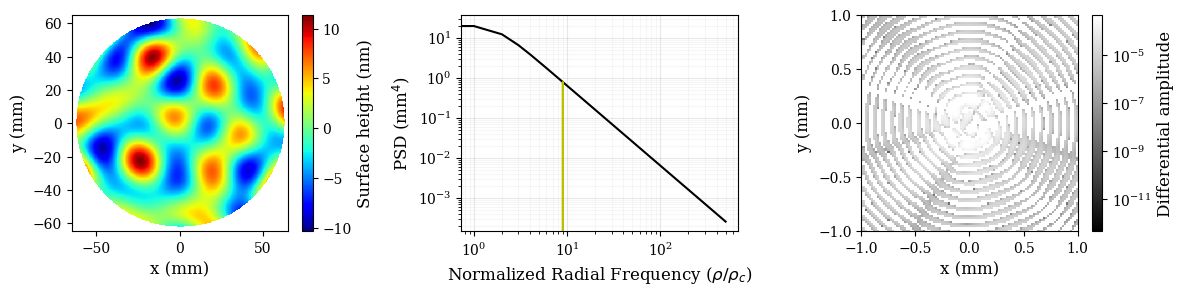}
        \label{fig:bandpass_2}
    \end{subfigure}\vspace{1cm}
    \\
    \begin{subfigure}[t]{\textwidth}
        \centering
        \includegraphics[width=\textwidth]{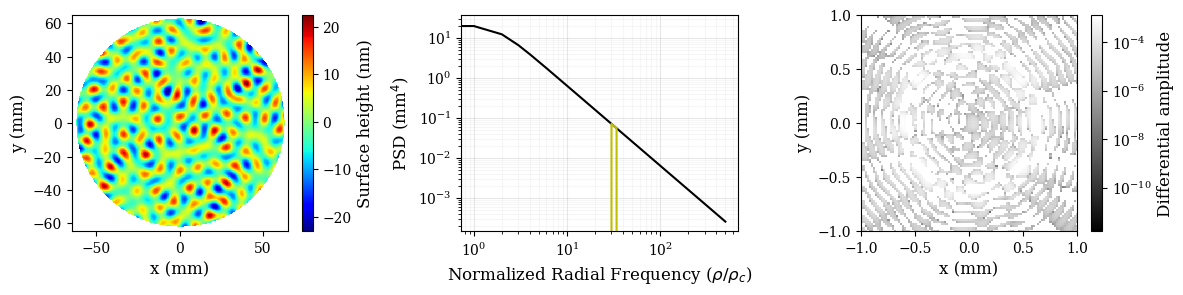}
        \label{fig:bandpass_3}
    \end{subfigure}
    \caption{Three instances of the WFE filtered by a small spatial frequency bandpass. As the mean spatial frequency of the bandpass filter increases, some of the power is observed to move radially outwards. The left panel shows the spatially filtered wavefront error on the ADC surface. The middle panel shows the part of the PSD curve that is used for the filtered WFE (orange line) and the original full PSD curve (black line). Finally, the right panel shows the differential amplitude between the unperturbed PSF and the PSF where the WFE in the first panel has been applied at the ADC.}
    \label{fig:bandpass_multifigure}
\end{figure}

 

\end{document}